\documentclass[showpacs,amssymb,twocolumn,aps]{revtex4}
\usepackage{amsmath}
\usepackage{graphicx} 
\usepackage{epsf}
\usepackage{graphics}
\usepackage[latin1]{inputenc}
\newcommand{\be}{\begin{equation}}
\newcommand{\ee}{\end{equation}}
\newcommand{\bea}{\begin{eqnarray}}
\newcommand{\eea}{\end{eqnarray}}

\begin{document}

\title{Non-Quadratic Gauge Fixing and Ghosts for Gauge Theories on the Hypersphere}

\author{F. T. Brandt$^{a}$  %\footnote{fbrandt@usp.br},
 and D. G. C. McKeon$^{b,c}$} %\footnote{dgmckeo2@uwo.ca} }
\affiliation{$^a$ Instituto de F\'{\i}sica,
Universidade de S\~ao Paulo,
S\~ao Paulo, SP 05315-970, Brazil}
\affiliation{$^b$ Department of Applied Mathematics, University of
Western Ontario, London, ON  N6A 5B7, Canada}
\affiliation{$^c$ Department of Mathematics and Computer Science, Algoma University, Sault St.Marie, ON P6A 2G4, Canada}

\begin{abstract}
It has been suggested that using a gauge fixing Lagrangian that is not quadratic in a gauge fixing 
condition is most appropriate for gauge theories formulated on a hypersphere. We reexamine the
appropriate ghost action that is to be associated with gauge fixing, applying a technique that has been
used for ensuring that the propagator for a massless spin-two field is transverse and traceless.
It is shown that this non-quadratic gauge fixing Lagrangian leads to two pair of complex Fermionic
ghosts and two Bosonic real ghosts.
\end{abstract}


\maketitle

It has been shown that the classical action for Yang-Mills theory on an $n$-dimensional hypersphere
\cite{Dirac:1935,Adler:1972,Jackiw:1976,Drummond:197989,Shore:1979121} is
\be\label{eq1}
S_{cl} = \int d\Omega \left[-\frac{1}{12} \left(F^i_{abc}\right)^2\right],
\ee
where $d\Omega$ is a surface element of a hypersurface of unit radius in this $n$-dimensional Euclidean
space with coordinates $\eta_a$ ($a = 1,\cdots, n$) and
\begin{eqnarray}\label{eq2}
F_{abc}^i &=&  L_{ab} A_c^i + f^{ijk} \eta_a A^j_b A_c^k + {\mbox{(\rm cyc. perm.)}}  
\nonumber \\
& & (L_{ab} = \eta_a \frac{\partial}{\partial \eta_b} - \eta_b \frac{\partial}{\partial \eta_a})   
\end{eqnarray}
This action is invariant under the two gauge transformations
\begin{subequations}\label{eq3}
\begin{eqnarray}\label{eq3a}
\delta_{\phi} A^i_ a &=& \eta_a \phi^i(\eta)  \\
\label{eq3b}
\delta_{\theta} A^i_ a &=& \eta_p L_{p a} \theta^i(\eta) + f^{ijk} A^j_a \theta^k(\eta). 
\end{eqnarray}
\end{subequations}
(These transformations are infinitesimal; this is adequate for our purpose.)

In the path-integral formulation of the generating functional, it has been pointed out 
\cite{Drummond:197989, Shore:1979121} that the most convenient gauge fixing Lagrangian is
\be\label{eq4}
{\cal L}_{gf} = \frac{1}{2\alpha} \left[\left(L_{ab}+\delta_{ab}\right)A^i_b\right]
\left[\left(L_{ac}-(n-2)\delta_{ac}\right)A^i_c\right]
\ee
The technique used in Ref \cite{Shore:1979121} to derive the ghost action associated with this gauge fixing does not seem to work for standard Yang-Mills theory in Euclidean space in the Lorenz-Feynman gauge (even though it possesses a BRST invariance \cite{McKeon:1991}); consequently we are motivated to reexamine how the ghost contribution to the effective action arises when one uses the non-quadratic gauge fixing Lagrangian of Eq. \eqref{eq4}. The approach of Refs. \cite{Brandt:2007td, Brandt:2009qi} can be used. (In the two references the transverse-traceless propagator for a spin-two field and the propagator for a spontaneously broken gauge theory are considered using non-quadratic gauge fixing.)

The technique we use is a generalization of 't Hooft's approach \cite{Hooft:1971fh} to derive the Faddeev-Popov ghost action \cite{Faddeev:1967fc}. We first insert the following constant factors
\begin{widetext}
\begin{subequations}\label{eq5}
\begin{eqnarray}\label{eq5a}
{\rm det}\left[\left(L_{ab} + \delta_{ab} \right)\eta_b \delta^{ij} \right] 
{\rm det}\left[\left(L_{ab} + \delta_{ab} \right)\left( \eta_p L_{p b} \delta^{ij} + f^{ip j} A^p_b\right)\right]   
\nonumber \\   \times
\int {\cal D} \phi_1  {\cal D} \theta_1 \delta\left[\left(L_{ab} + \delta_{ab}\right)
\left(A^i_b + \eta_b\phi^i_1 + \eta_p L_{p b} \theta^i_1 +  f^{ijk} A^j_b \theta^k_1\right) -p^i_a\right] 
\end{eqnarray}
and
\begin{eqnarray}\label{eq5b}
{\rm det}\left[\left(L_{ab} - (n-2) \delta_{ab} \right)\eta_b \delta^{ij} \right] 
{\rm det}\left[\left(L_{ab} - (n-2) \delta_{ab} \right)\left( \eta_p L_{p b} \delta^{ij} + f^{ip j} A^p_b\right)\right]   
\nonumber \\   \times
\int {\cal D} \phi_2  {\cal D} \theta_2 \delta\left[\left(L_{ab} -(n-2) \delta_{ab}\right)
\left(A^i_b + \eta_b\phi^i_2 + \eta_p L_{p b} \theta^i_2 + f^{ijk} A^j_b \theta^k_2\right) -q^i_a\right] 
\end{eqnarray}
\end{subequations}
\end{widetext}
into the path integral
\be\label{eq6}
Z[J_a^i] =\int{\cal D} A^i_a \exp{\left\{\int d\Omega\left[-\frac{1}{12} \left(F^i_{abc}\right)^2
+ J_a^i A_a^i\right] \right\}}
\ee
We follow this with an insertion of the constant factor
\be\label{eq7}
\int {\cal D} p^i_a {\cal D} q^i_a\exp{\left(\frac{1}{2\alpha}\int d\Omega p_a^i g_ a^i\right)}
\ee
into $Z[J]$. If now we perform the gauge transformation
\be\label{eq8}
A_a^i \rightarrow A_a^i - \delta_{\phi_1} A^i_a - \delta_{\theta_1} A^i_a
\ee
in the resulting expression for $Z[J]$ and then do the integration over $p_a^i$ and $q_a^i$, we end up with
\begin{widetext}
\begin{eqnarray}\label{eq9}
Z[J_a^i] &=&\int{\cal D} A^i_a \int {\cal D} \phi_1  {\cal D} \theta_1 \int {\cal D} \phi  {\cal D} \theta 
{\rm det}\left[\left(L_{ab} + \delta_{ab} \right)\eta_b \delta^{ij} \right] 
{\rm det}\left[\left(L_{ab} - (n-2) \delta_{ab} \right)\eta_b \delta^{ij} \right] 
\nonumber \\ &\times&
{\rm det}\left[\left(L_{ab} + \delta_{ab} \right)\left( \eta_p L_{p b} \delta^{ij} + f^{ip j} A^p_b\right)\right]   
{\rm det}\left[\left(L_{ab} - (n-2) \delta_{ab} \right)\left( \eta_p L_{p b} \delta^{ij} 
+ f^{ip j} A^p_b\right)\right]   
\nonumber \\ &\times&
\exp\;\int d\Omega\left[-\frac{1}{12} \left(F^i_{abc}\right)^2
+\frac{1}{2\alpha}\left[\left(L_{ab} + \delta_{ab}\right)A_b^i\right] 
\right. \nonumber \\ &\times& 
\left. 
\left[\left(L_{ac} - (n-2) \delta_{ac} \right)\left(A_c^i+\eta_c\phi^i+\eta_p L_{p c}\theta^i
+f^{ijk} A^j_c \theta^k\right) \right] +  J_a^i A_a^i\right]
;\;\; \left(\phi^i = \phi_2^i-\phi_1^i, \;\; \theta^i = \theta_2^i-\theta_1^i\right) .
\end{eqnarray}
\end{widetext}
We have used the fact the the functional determinants in Eqs. \eqref{eq5a} and \eqref{eq5b} are gauge invariant.

Each of the functional determinants in Eq. \eqref{eq9} can be exponentiated using the relation
\be\label{eq10}
\int {\cal D} c_a{\cal D} \bar{c}_a {\rm exp}\left[\bar{c}_a M_{ab} c_b \right] = \det\, M,
\ee
where $c_a$ and $\bar{c}_a$ are complex fermionic ``ghost'' fields.
Consequently there are two pair of complex Fermionic ``ghost''  fields arising from the functional determinants appearing in Eq. \eqref{eq9}; in addition there are the Bosonic ``ghost''fields $\phi^i$ and $\theta^i$. The functional integrals over $\phi_1$ and $\theta_1$ serve only to rescale the normalization function for $Z[J_a]$. The gauge fixing Lagrangian of Eq. \eqref{eq4} has thus served to complicate the propagator for $A^i_a$ because of the nature of these terms that occur in the exponential appearing in Eq. \eqref{eq9}  which are quadratic in the fields $A_a^i$, $\phi^i$ and $\theta^i$. By a ``completing the square''  operation, these quadratic terms can be made diagonal in the fields
$A^i_a$, $\theta^i$ and $\phi^i$, thereby retaining the vector propagator of Refs. \cite{Drummond:197989,Shore:1979121}, though at the expense of having more complicated interaction terms.

\begin{acknowledgments}
F. T. Brandt would like to thank CNPq for financial support and
the hospitality of the Department of Applied Mathematics of the University of Western Ontario, Canada. 
D. G. C. McKeon would like to thank R. MacLeod for a helpful suggestion, the Instituto de 
Física da Universiade de São Paulo, for the hospitality, and FAPESP, Brazil.
\end{acknowledgments}

%
%
%\bibliographystyle{apsrev.bst}
%\bibliography{all_new}

\end{document}